\documentclass[12pt]{article}

\usepackage{newtxtext,newtxmath}

\usepackage{graphicx}

\usepackage[letterpaper,margin=1in]{geometry}

\renewenvironment{abstract}
	{\quotation}
	{\endquotation}

\date{}

\makeatletter
\renewcommand{\fnum@figure}{\textbf{Figure \thefigure}}
\renewcommand{\fnum@table}{\textbf{Table \thetable}}
\makeatother

\usepackage{scicite}

\usepackage{url}
\usepackage{comment}

\def\scititle{
	Quantum Sensing of Broadband Spin Dynamics and Magnon Transport in Antiferromagnets
}
\title{\bfseries \boldmath \scititle}

\author{
	Alex~L.~Melendez$^{1\dagger}$,
    Shekhar~Das$^{1\dagger}$,
	Francisco~Ayala~Rodriguez$^{1}$,\and
    I-Hsuan~Kao$^{2}$,
    Wenhao~Liu$^{3}$,
	Archibald~J.~Williams$^{4}$,
    Bing~Lv$^{3}$,\and
    Joshua~Goldberger$^{4}$,
    Shubhayu~Chatterjee$^{2}$,
    Simranjeet~Singh$^{2 \ast}$,
    P.~Chris~Hammel$^{1 \ast}$\and
	\small$^{1}$Department of Physics, The Ohio State University, Columbus, OH 43210, USA.\and
	\small$^{2}$Department of Physics, Carnegie Mellon University, Pittsburgh, PA 15213, USA.\and
    \small$^{3}$Department of Physics, The University of Texas at Dallas, Richardson, TX 75080, USA.\and
    \small$^{4}$Department of Chemistry and Biochemistry, The Ohio State University, Columbus, OH 43210, USA.\and
	\small$^\ast$Corresponding authors. Email: simranjs@andrew.cmu.edu and hammel.7@osu.edu\and 
	\small$^\dagger$These authors contributed equally to this work.
}

\begin{document} 

\maketitle
    
\begin{abstract} \bfseries \boldmath
Optical detection of magnetic resonance using quantum spin sensors (QSS) provides a spatially local and sensitive technique to probe spin dynamics in magnets. However, its utility as a probe of antiferromagnetic resonance (AFMR) remains an open question. We report the first experimental demonstration of optically detected AFMR in layered van der Waals antiferromagnets (AF) up to frequencies of 24 GHz. We leverage QSS spin relaxation due to low-frequency magnetic field fluctuations arising from collective dynamics of magnons excited by the uniform AFMR mode. First, through AFMR spectroscopy we characterize the intrinsic exchange fields and magnetic anisotropies of the AF. Second, using the localized sensitivity of the QSS we demonstrate magnon transport over tens of micrometers. Finally, we find that optical detection efficiency increases with increasing frequency. This showcases the dual capabilities of QSS as detectors of high frequency magnetization dynamics and magnon transport, paving the way for understanding and controlling the magnetism of antiferromagnets.
\end{abstract}

\subsection*{Introduction}
\noindent
Magnetic resonance is a spectroscopically precise tool capable of probing spin interactions and revealing detailed aspects of magnetic dynamics.  
The understanding of spin dynamics in antiferromagnets (AFMs), still in its infancy, is essential to realize conceptualized spintronics devices for generation, transmission, and detection of high frequency signals \cite{BaltzRevModPhys,SmejkalNatPhysTopoAFSpintronics}. 
In recent years van der Waals (vdW) AFMs such as CrCl$_3$ and CrSBr \cite{Huang1,Huang,McGuire,lee2024spinorbitcouplingcontrolledtwodimensional,Luqiao,DanRalphInductiveAFMRCrSBr,CrSBrRoss,CrSBrBingLv,BingLvNatComms} have emerged as a promising platform for studying spin dynamics.
Compared to traditional AFMs such as metal oxides or fluorides, the smaller AFM exchange in vdW AFMs allows experimentation to probe the fundamentals of antiferromagnetic spin dynamics at relatively lower frequencies.
This calls for new experimental tools that are able to provide a sensitive and spatially localized measurement of spin dynamics of AFMs over a broad frequency spectrum. 
Developing and utilizing such tools will be central to understanding emergent dynamical phenomena in AFMs for spintronic applications. 

In recent years color defects in solids, such as NV$^{-}$ centers in diamond, have emerged as sensitive optical probes enabling optical detection of ferromagnetic resonance (FMR) and hence magnetic dynamics, through their coupling to the fluctuating stray magnetic fields generated by magnons in ferromagnets \cite{Wolfe,vanderSarYacobyProbingSpinWaves,Chunhui_chemicalpotential}.
The NV$^{-}$ center is an atomic-scale quantum spin sensor (QSS) whose spin-dependent photoluminescence locally probes static and dynamic stray fields \cite{Degen_AnnualReviewPChem,Wrachtrup_2006_Review,JacquesAFMRnatcomm2020}.
In this role, its sensitivity to magnetization dynamics arises from the enhancement of the QSS spin relaxation rate by the fluctuating fields emanating from the magnon gas in the sample, specifically those magnetic field fluctuations matching the QSS ESR frequency $\omega_{NV}$.   
However, an experimental demonstration of the utility of the NV$^{-}$ center for optical detection of magnetic resonance in antiferromagnets (AFM), \emph{i.e.,} ODAFMR, wherein the characteristic resonant frequencies can range from a few GHz to 100’s GHz, is critically missing. 

Here we provide the first experimental demonstration of the detection of ODAFMR, using NV$^{-}$ centers as QSSs, in two vdW antiferromagnets: CrCl$_3$ and CrSBr.
Our main results are as follows.
First, we illustrate the effectiveness of the QSS relaxometric method through temperature and magnetic field dependent optical measurements of the magnetic resonance modes of CrCl$_{3}$ and CrSBr, and extract the intrinsic exchange fields and magnetic anisotropies from the ODAFMR spectra. Our results are consistent with previous measurements that employ conventional inductive methods \cite{Luqiao,DanRalphInductiveAFMRCrSBr}. 
Next, we show that the effectiveness of this technique improves with increasing frequency up to frequencies exceeding $\omega_{NV}$ by almost an order of magnitude, engendering optimism that this approach can be extended to higher frequencies paving the way to probe spin dynamics in a wide variety of AFMs using QSSs.  
Finally, we exploit the spatial localization of these field fluctuations to a length scale set by the magnon wavelength which, in tandem with the highly localized sensitivity of the nano-scale QSS, reveals magnon transport over length scales of tens of micrometers.


\subsection*{Results}
\subsubsection*{Materials Studied}

\begin{figure}
    \centering
    \includegraphics[width=0.75\textwidth]{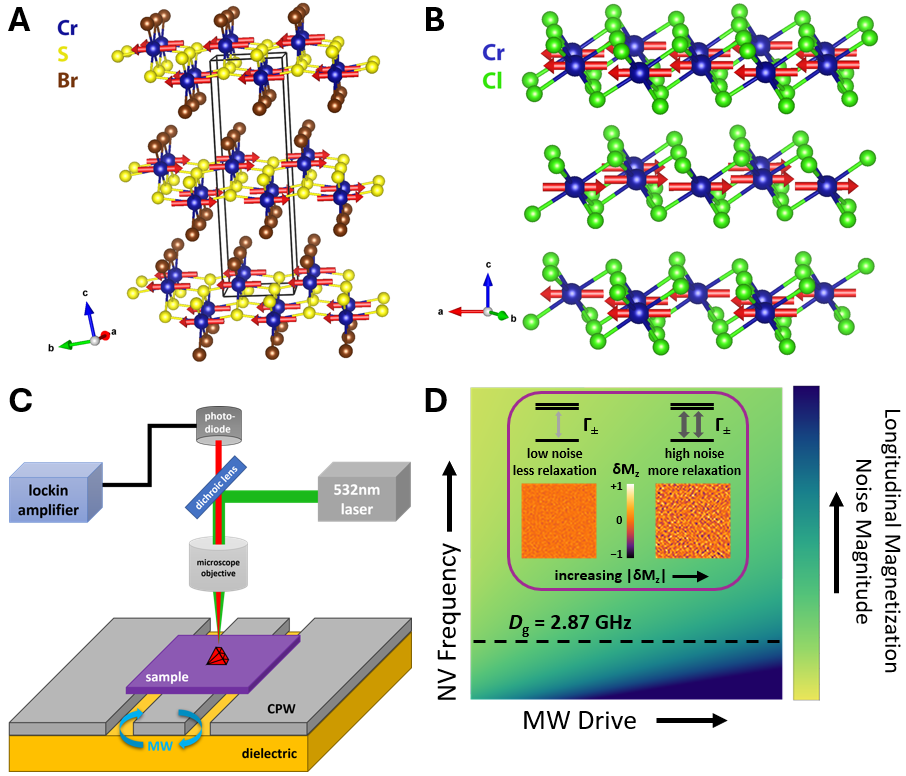}
    \caption{ 
    {\bf Overview of optical detection of high frequency magnetic dynamics in vdW antiferromagnets enabled by magnetic field noise arising from diffusing magnons.}
    Panels (A) and (B) provide a schematic side view of the crystal and spin structures %
    of CrSBr (A) and CrCl$_{3}$ (B), with spins indicated by red arrows. 
    Panel (C) is a schematic of the setup for optical detection of antiferromagnetic resonance. A sample flake is exfoliated and transferred onto a microwave coplanar waveguide. Nanodiamonds containing NV$^{-}$ centers are dropcast onto the sample surface and photoluminescence is measured under continuous laser excitation as a function of microwave frequency and applied in-plane magnetic field.
    Panel (D) schematically depicts  the magnetic noise magnitude as a function of the microwave drive power (horizontal axis) and the QSS frequency (vertical axis). Driving AFMR increases the magnon population in the sample, in turn increasing low-frequency magnetic noise even when the single-magnon gap is much larger than the QSS frequency as described in main text and SI. 
    The inset to panel (D) depicts the enhancement of the QSS relaxation rate due to the collective diffusive dynamics of the magnon gas at larger magnon-density.   
    }
    \label{fig:1}
\end{figure}

We studied two vdW AFMs: CrCl$_{3}$ and CrSBr (%
whose N\'eel temperatures $T_N$
are 14 and 132\,K, respectively) with dimensions approximately $1\,\text{mm} \times 1 \, \text{mm}\times 50 \, \mu$m thick.  These collinear antiferromagnets consist of two oppositely-oriented but equally magnetized spin sublattices that can be viewed as interleaved ferromagnetic sublattices (see Figs.~\ref{fig:1}A and B). Static stray fields vanish except at inhomogeneities such as grain boundaries, step edges or domain walls. However finite temperature and microwave magnetic field drive of the uniform AFMR mode excites magnons causing sublattice precession that creates a small dynamic net magnetization due to canting (see insets to Fig.~\ref{fig:2}) \cite{Vaidya_2020,RezendeAntiferromagneticMagnons} and hence stray fields that extend above the surface over distances set by the inverse of $k$. \cite{PurserNVseparation}

The samples were placed on a copper coplanar waveguide (CPW) and nanodiamonds containing NV$^{-}$ centers were dropcast onto the top surface as schematically depicted in Fig.~\ref{fig:1}C.  A static magnetic field is applied parallel to the sample plane and the CPW generates microwave magnetic fields. The CrCl$_{3}$ ODAFMR was measured with the static magnetic field applied in-plane but perpendicular to the CPW center conductor in a parallel-pumping geometry. Given the small in-plane anisotropy of CrCl$_{3}$, the sublattices maintain a spin-flop configuration resulting in optical and acoustic AFMR modes \cite{Luqiao}. CrSBr exhibits a larger in-plane anisotropy, that is, its anisotropy is bi-axial \cite{DanRalphInductiveAFMRCrSBr}; 
the CrSBr was oriented with the easy-anisotropy ($b$) axis oriented parallel to the applied field (see Fig.~\ref{fig:S4} and related discussion in Supplemental Information (SI) for discussion of magnetic configurations).
The excitation of the uniform AFMR mode thus excites modes that are hybrids of the right-handed and left-handed uniaxial AFMR modes \cite{DanRalphInductiveAFMRCrSBr}. The insets to Fig.~\ref{fig:2} depict these modes in spin-space.

We measure the magnetic field dependence of the change in photoluminescence in response to excitation of AFMR at various microwave frequencies using a lock-in amplifier while modulating the amplitude of the applied microwave field (Fig.\ref{fig:2}).


\subsubsection*{Principles of ODAFMR}

The relative occupation of the NV$^{-}$ spin states can be manipulated by microwave magnetic fields matching the electron spin resonance (ESR) frequency. Coherent irradiation resonantly drives $\Delta m = \pm 1$ transitions between NV$^{-}$ spin states (ESR) while stochastic fluctuating fields relax a spin system that is out of thermal equilibrium with the thermal reservoir to which it is coupled by these fluctuating fields---the magnon gas in this case---as is the case for NV$^{-}$ spins when they are hyperpolarized by laser irradiation.  The magnetic Hamiltonian of the NV$^{-}$ spin in an applied field $\boldsymbol{B}$ is given by \cite{Maletinsky_Jacques}
\begin{align}
\frac{\hat{\mathcal{H}}_{\text{NV}}}{h} 
    = D_{g}\left(\hat{S}_z^2-\dfrac{S(S+1)}{3}\right)  
    + E_{g} \left( \hat{S}_x^2 - \hat{S}_y^2 \right) 
    + \frac{\tilde{\gamma}}{2\pi} \boldsymbol{B} \cdot \hat{\boldsymbol{S}},
\end{align}
where $h$ is Planck's constant, $\tilde{\gamma}/2\pi=2.8$\,MHz/G is the electron gyromagnetic ratio, $D_g=2.87$\,GHz is the ground state zero-field splitting (ZFS), $E_g$ is the off-axis splitting, $S=1$ is the NV$^{-}$ spin magnitude, and $\hat{\boldsymbol{S}}=\left[\hat{S}_x,\,\hat{S}_y,\,\hat{S}_z\right]^T$ is the vector of spin-1 Pauli matrices. Thus, stochastic electromagnetic fluctuations matching the $m_s=0 \,\leftrightarrow\, m_s=\pm1$ transition frequencies will relax the NV$^{-}$ spin polarization, enhancing the population of the less radiative $m_s = \pm 1$ states and reducing photoluminescence. This is the principle of NV$^{-}$ center relaxometry as a method for optically detecting magnetic dynamics arising from the excitations of AFMR.

The fluctuating magnetic fields that relax the NV$^{-}$ spin result from the diffusive collective dynamics of the magnon gas in the proximate vdW magnet.  
Thus, even when the single-magnon AFMR dispersion has a large spectral gap and hence does not contribute directly to magnetic noise at the NV$^{-}$ ESR frequency $\omega_{NV}$, magnon-magnon processes generate electromagnetic fluctuations whose spectrum extends to very low frequency \cite{FlebusTserkovnyak, Brendan, Chunhui_chemicalpotential, Chunhui_hematite_ambient,CRD2018} as schematically illustrated in Fig.~\ref{fig:1}D.  In the absence of microwave drive the density of magnons is described by a thermal magnon population.
When the uniform ($\boldsymbol{k}=0$) AFMR mode is resonantly excited by monochromatic microwaves, the resultant enhanced spin-density decays into finite-momentum ($\boldsymbol{k} \neq 0$) magnons through non-linear interactions in the Hamiltonian (see SI for details). 
This increased population of magnons subsequently generates strong low-frequency magnetic field fluctuations that match the NV$^{-}$ spin-transition frequency, thereby enhancing the NV$^{-}$ relaxation rate.

To formalize the above intuition, we note that NV$^{-}$ relaxation rate resulting from stochastic field fluctuations is given by the spectral density $\Gamma (\omega) $ of these stray magnetic fields $\hat{B}^{\pm } $ evaluated at $\omega_{\rm NV}$: 
\begin{align}\label{spectraldensity}
    \Gamma (\omega_{NV}) =  
    \frac{\gamma^2}{2} 
    \int\langle\hat{B}^+(0)\,\hat{B}^-(t)\rangle e^{- i \omega_{NV} t} \text{d}t 
\end{align}
where  $\gamma$ is the NV$^{-}$ gyromagnetic ratio. 
The larger number of finite wavevector magnons that are excited indirectly via microwave driving of the uniform AFMR mode enhances the amplitudes of the stray fields $\hat{B}(t)$ and hence $\Gamma(\omega_{NV})$.
More explicitly, the stray-field correlation function can be expressed in terms of the longitudinal spin-fluctuations $C_{\parallel}(\boldsymbol{k},\omega_{NV})$ of the magnon gas, as 
\begin{align}
\Gamma(\omega_{NV}) \propto \int d^2k \, F_d(k) C_{\parallel}(\boldsymbol{k},\omega_{NV}),
\end{align}
where $F_d(k) = k^2 e^{-2kd}$ is a momentum-filter function that describes the propagation of the stray-field fluctuations from the source (magnet) to the sensor (NV$^{-}$) which is at a distance $d$. \cite{CRD2018,FlebusTserkovnyak,Machado2023}.
Assuming that the magnon-density $n$ is approximately conserved in the steady state, the magnetization density is also conserved and the longitudinal spin-correlation function $C_{\parallel}(\boldsymbol{k},\omega_{NV})$ takes a diffusive form, with a diffusion constant $D_s$ that scales inversely with $n$. 
In this limit, $\Gamma(\omega_{NV}) \propto D_s^{-2} \propto n^2$ (see SI for a derivation), implying that the NV$^{-}$ relaxation rate is enhanced due to a larger density of magnons. 
Consequently, the uniform AFMR mode may be characterized directly via NV$^{-}$ relaxometry through microwave driving, in spite of its spectral gap being much larger than $\omega_{NV}$, as we show below.


\subsubsection*{ODAFMR Measurements}

\begin{figure}
    \centering
    \includegraphics[width=0.75\textwidth]{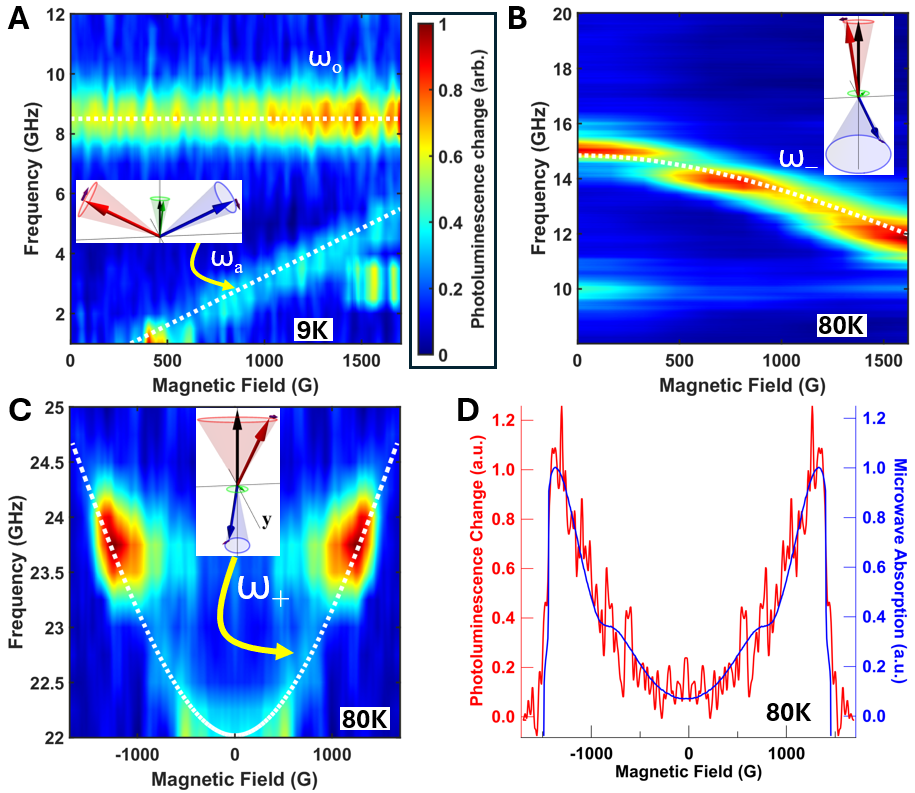}
    \caption{{ \bf Optically detected antiferromagnetic resonance spectra up to 24 GHz.}      
    Panel (A):  Optically-detected CrCl$_{3}$ AFMR dispersion showing the optical mode $\omega_o$ and acoustic mode $\omega_a$; the white dashed lines show fits using Eq.~\ref{modes}. 
    The insets to panels (A)-(C) depict the motion of the AF sublattice magnetizations in spin space in response to AFMR excitation. 
    Sublattice magnetizations $\boldsymbol{M}_A$ and $\boldsymbol{M}_B$ are shown as red and blue arrows respectively, the net magnetization in green, and applied field in black.
    The cones trace out the motion of the magnetization over a cycle, with the solid ellipses tracing the path of the arrowhead.
    The small purple arrows indicate the direction of precession of each sublattice. 
    The equilibrium orientations about which the sublattice magnetizations precess are shown in Figs.\,\ref{fig:S4}A and \ref{fig:S4}B.
    Panels (B)-(D) present the optically measured dependencies of the CrSBr AFMR mode frequencies on magnetic field: 
    Panel (B) shows the low frequency mode $\omega_-$ [see Eq.~\ref{eq:CrSBr_modes} for $\omega_- (H)$ and $\omega_+ (H)$], 
    while Panel (C) shows the high frequency mode $\omega_+$. The internal fields corresponding to the white dashed lines are presented in Fig.~\ref{fig:3}D. 
    Panel (D) shows linecuts through the data in panel (C) at 23.75\,GHz comparing the optically detected CrSBr $\omega_+$ AFMR mode (red) with simultaneously measured inductive AFMR (blue).
    }
    \label{fig:2}
\end{figure}

We demonstrate the efficacy of optically-detected AFMR spectroscopy using NV$^{-}$ relaxometry in two widely studied vdW AFs:  CrCl$_3$ and CrSBr.  For this novel approach we first confirm in both cases that spectra obtained are in agreement with conventionally obtained microwave measurements and enable spectroscopic determination of key magnetic parameters such as exchange interactions and anisotropies. Temperature-dependent CrCl$_3$ ODAFMR spectra---the dependence of PL reduction on microwave frequency and static in-plane magnetic field at 9 K (Fig.~\ref{fig:2}A).  We then study CrSBr at five temperatures cooling from 120 K---where AFMR vanishes at a transition to what we believe is a spin-flop phase---down to 80 K where the AFMR frequency rises to 24 GHz.

We observe two CrCl$_{3}$ AFMR modes (see Fig.~\ref{fig:2}A) corresponding to the acoustic (lower) and optical (upper) AFMR modes, at frequencies $\omega_\text{a}$ and $\omega_\text{o}$ respectively \cite{Luqiao}:
\begin{align}
    \omega_\text{a}(H) = \mu_0\gamma\sqrt{\frac{2 H_E + M_S}{2H_E}}\,H,
    && \text{and} &&
    \omega_\text{o}(H) = \mu_0\gamma\sqrt{2 H_E M_S \left( 1-\frac{H^2}{4 H_E^2} \right)} 
    \label{modes}
\end{align}
where $H_E$ is the exchange field, $M_S$ is the saturation magnetization, and $\gamma$ is the gyromagnetic ratio of CrCl$_{3}$. In the field range accessible to our setup, the optical mode $\omega_\text{o}$ appears approximately constant at about $\sim 8.5$\,GHz. Consistent with previous 
AFMR
measurements 
obtained using conventional microwave absorption techniques  \cite{Luqiao},
this mode is relatively broad in frequency. Fitting our optical data yields estimates of the exchange field and saturation magnetization of CrCl$_{3}$: $\mu_0 H_E = (2.64 \pm 0.17)$\,kG and $\mu_0 M_S = (1.75 \pm 0.11)$\,kG at 9\,K assuming $\gamma/2\pi=2.8$ MHz/G. Simultaneous inductive AFMR measurements (see Fig.~\ref{fig:S2} A, Supplemental Information), manifesting as a reduction in the microwave transmission $S_{12}$ give similar values of $\mu_0H_E=(2.58\pm0.22)$\,kG and $\mu_0M_S=(1.78\pm0.15)$\,kG.

Turning to CrSBr, another easy-plane AFM with hard and soft magnetic axes in the easy plane, we present ODAFMR spectra obtained over a wider range of temperatures to demonstrate the effectiveness of ODAFMR as a local magnetic spectroscopy tool to determine magnetic parameters; these are shown in Fig.~\ref{fig:3}.
Given the biaxial anisotropy in CrSBr (the hard $c$-axis is perpendicular to the intermediate $a$-axis), the AFMR modes $\omega_\pm$ with $\boldsymbol {H}$ parallel to the easy $b$-axis are hybridizations of the uniaxial right-handed (RH) and left-handed (LH) modes \cite{DanRalphInductiveAFMRCrSBr,RezendeAntiferromagneticMagnons}:
\begin{align}\label{eq:CrSBr_modes}
    &\frac{\omega_\pm(H)}{\mu_0\gamma} = 
    & \sqrt{ H^2 + H_a ( H_E + H_c ) + H_E H_c 
    \pm \sqrt{H^2 ( H_a + H_c ) ( H_a + 4 H_E + H_c ) + H_E^2 (H_a - H_c )^2 }},\nonumber \\
\end{align}
where the subscripts $\pm$ refer to higher and lower frequency and are unrelated to the superscripts used in Eq.\,\eqref{spectraldensity}. Here $H_c$ and $H_a$ are the hard and intermediate axis anisotropies respectively. Fig.~\ref{fig:2}B shows ODAFMR of the lower mode $\omega_-$ at 80\,K. To enhance the magnon density, data were typically taken at input microwave power of 30\,dBm, a power sufficient that an asymmetric lineshape characteristic of non-linear response \cite{Brendan} is observed (see Supplemental Information).  However transmission losses substantially reduce the power delivered to the CPW: at the highest frequencies the power applied to the CPW is reduced by approximately two orders of magnitude (see inset, Fig.~\ref{fig:4}B). The microwave drive and the laser increase the sample temperature above ambient by less than 5\,K (see Supplemental Information).

The CrSBr ODAFMR spectra obtained at 90, 100, 110 and 120\,K are shown in Fig.~\ref{fig:3}. Values of the effective fields obtained from the optical data are shown in Fig.~\ref{fig:3}D: the effective fields $H_E$, $H_c$ and $H_a$ and hence both modes $\omega_\pm$ decrease with increasing temperature as previously observed \cite{DanRalphInductiveAFMRCrSBr}. At 120\,K a feature arising from either a spin-flop or ferromagnetic mode appears as temperature approaches $T_N$. Similar features are also observable around 1500\,G at 110K.

\subsection*{Discussion}
Based on our results, we discuss two salient observations that highlight the efficacy of optically detected relaxometry for understanding spin dynamics in AFs.
First, we show that the local sensitivity of the QSS can be leveraged to demonstrate transport of AF magnons across the $50 \, \mu$m thickness of the CrSBr crystal, 
revealing long-range magnon transport indicative of a magnon mean-free-path comparable to the crystal thickness (tens of microns).
%
Our observation is consistent with recent magneto-optical observations of long-ranged spin-transport in CrSBr 
{\cite{YJBaeNatureCoherentMagnonsCrSBr,Orenstein2024dipolar}}.
%
Second, we demonstrate relaxometric detection of ODAFMR detection extending to 24\,GHz, the highest frequency reported to date and eight times the QSS resonance frequency.
We further find that, at constant CPW-generated microwave magnetic field intensity, the efficiency of ODAFMR detection substantially \emph{increases} with increasing frequency, thereby enabling broadband detection of spin dynamics.

To show long-range magnon transport, we begin by noting that the sensitivity of the QSS to the fluctuating stray magnetic fields emanated by the magnon gas in the CrSBr sample is highly local. 
Due to the evanescent nature of the magnon-generated electromagnetic wave, the sensitivity of the QSS to the collective magnetic mode decreases exponentially with the mode wavevector $k$ as described by the filter function $F_d(k) = k^2 e^{-2kd}$ \cite{Chunhui_chemicalpotential,FlebusTserkovnyak,Machado2023} mentioned earlier, where $d$ is the separation of the QSS source of the stray field.  
Consequently, this suppressed coupling to magnetic modes at the bottom of the sample (50 $\mu$m thick) minimizes the contribution of magnons located there to QSS relaxation. 

At the same time, 
the current density in the CPW center conductor is confined to its edges due to a combination of the skin effect (skin depth $\delta _s \sim 1\,\mu m$ at 10\,GHz) and charge bunching (also known as the proximity effect) \cite{FoundationsMicrostripDesign_ch7,HighFrequencyCPWchiang2023}. This edge-confinement of current filaments attenuates the microwave fields at the top of the sample 50 $\mu m$ above the CPW.  This decreased microwave field strength is confirmed by our NV ESR measurements that revealed ESR signals much weaker than those from NV$^{-}$ centers in nanodiamonds deposited on the CPW.  These microwave fields are weaker still at AFMR frequencies (relative to microwave excitation power---see below): too weak to excite AFMR at the top of the sample sufficiently to account for the rapid QSS relaxation rates we observe.
%
This implies that magnons generated at the bottom of the crystal readily diffuse tens of microns to the top of the sample where they couple strongly %
the QSS thus enhancing its relaxation rate.  

To quantify the above observation, we compare the strength of the coherently driven ESR signal to the relaxometric signal (remembering that in both cases the fluctuating fields match the QSS resonance frequency, but the fields that relax the QSS originate from excitation of AFMR at much higher frequencies). 
Because these signals grow linearly with microwave power, we compare the strength of ESR---a measure of the square of the microwave fields generated by the CPW 50 $\mu$m below---vs.\ AFMR signals---a measure of magnon-generated fields---normalized by the microwave power delivered to the CPW at the relevant drive frequencies [Fig.~\ref{fig:4}A]. 
Due to 
losses in the CPW and associated transmission lines%
, the power delivered at the higher AFMR frequencies (up to 24\,GHz) is reduced by up to two orders of magnitude (see the inset to Fig.~\ref{fig:4}B). 
Nevertheless, a comparison of power-normalized signals in Fig.~\ref{fig:4} shows that the signal generated by exciting the  higher frequency (21.7\,GHz) AFMR mode actually \emph{exceeds the directly driven} NV$^{-}$ ESR by over an order of magnitude.  
This demonstrates remarkably strong fluctuating fields arising from a high density of finite-wavevector magnons at the top surface of the thick crystal.  
%
The CPW-generated microwave magnetic fields at AFMR frequencies are even weaker at the top surface than at ESR frequencies due to the frequency dependence of charge bunching.  This further reduces the ability of the distant CPW to drive AFMR sufficiently to generate magnon densities required for this strong relaxometric signal. Instead they must arise from magnons generated adjacent to the CPW that then
diffuse to the top surface of the crystal and accumulate there. 
This observation of magnon transport over tens of microns to the NV$^{-}$ detectors atop the crystal enhances the promise of AF magnonics for spintronic applications.  

Our second observation emphasizes the potential relaxometry holds for application of spatially localized AFMR to higher frequencies still:  the effectiveness of this relaxometric detection of AFMR by QSS increases with increasing frequency up to the highest frequencies we measured. 
Fig.~\ref{fig:4}B shows this frequency dependence:  The power-normalized signal at 21.7\,GHz exceeds that at 15.1\,GHz by a factor of 4.4.  
Future studies will seek to clarify the mechanism for this.


In summary, we show that quantum spin sensors are effective local optical probes of antiferromagnetic dynamics in the vdW antiferromagnets CrCl$_{3}$ and CrSBr. 
Specifically, we  characterize magnetic exchange fields and anisotropies using optically detected AFMR, and illuminate underlying magnetization dynamics and excitations. 
The localized sensitivity of the sensors to fluctuating stray fields generated by the AF magnons reveals magnon transport over tens of micrometers in these materials.
Studies of CrSBr were performed at frequencies up to 24\,GHz, eight times the NV$^{-}$ ESR frequency and  AFMR detection efficiency was found to increase with AFMR frequency.
By exploiting non-equilibrium magnon dynamics to infer high frequency spectroscopic features of AFMs, our work adds a new dimension to existing and concurrent studies of equilibrium low-energy spin-dynamics in 2D magnets using spin sensors \cite{Chunhui_hematite_ambient,huang2023layer,Huang2023RevealingMicroscope,Tschudin2023NanoscaleCrSBr,Xue2024SignaturesFerromagnet,ziffer2024quantum}.
Further, our observations of long-range magnon transport indicate the potential for 2D AFMs to excel in high frequency spintronic applications.

Magnetic resonance is a proven tool for spectroscopic characterization of interactions and dynamics in magnets.  While conventional inductively detected magnetic resonance techniques (microwave power absorption) lack the sensitivity and spatial localization needed for spatially localized studies, this work demonstrates that QSS-based ODAFMR overcomes this barrier to local studies of AFMs.   
Another promising avenue for expanding the scope of this technique lies in the application of the boron-vacancy  center embedded in hexagonal boron nitride \cite{Das.hBN.PhysRevLett.133.166704} whose spin-dependent optical properties are similar to those of the NV$^{-}$ center \cite{gong2024isotope}, but offers the advantage that it can be seamlessly integrated with a variety of 2D magnetic systems. 

\subsection*{Materials and Methods}
\subsubsection*{Materials}
CrCl$_3$ crystals (Fig.\,\ref{fig:S1}A) obtained from J.~Goldberger were grown using a typical chemical vapor transport method. Anhydrous CrCl$_3$ (99.99\%, Alfa Aesar) was weighed out in an Ar-filled glovebox and sealed in an evacuated quartz ampoule (270 mm length, 15 mm inner diameter, 1 mm wall thickness). The ampoule was heated in a three-zone furnace for 18-72 h.  The feed zone, which contained the reactant mixture, was kept at 650-700$^{\circ}$\,C, while the crystals grew in the growth zone at 500-550$^{\circ}$\,C. The resulting crystalline platelets had a purple color. Purity was confirmed by powder X-ray diffraction. The high quality CrSBr single crystals (Fig.\,\ref{fig:S1}B) obtained from B.~Lv were synthesized through a direct solid-vapor reaction in a box furnace, as described in detail in reference \cite{CrSBrBingLv}.
Both samples were exfoliated to a thickness of approximately 50\,$\mu$m.

\subsubsection*{Experimental Configuration}
Experiments were performed using a custom copper coplanar waveguide (CPW) with a center conductor width of 330\,$\mu$m, thickness 35\,$\mu$m and with the dielectric FR-4 from Advanced Circuits. 
The samples were placed on the center line of the microwave CPW.
Nanodiamonds containing NV$^{-}$ centers were then dropcast onto the surfaces of the samples resulting in a non-uniform distribution. 
Room temperature optically detected NV$^{-}$  Electron Spin Resonance (ESR) verified their presence on the surface of the CrCl$_{3}$ (Fig.\,\ref{fig:S1}C) and CrSBr (Fig.\,\ref{fig:S1}D). 
As expected no ODAFMR signal is visible for either sample at room temperature.

\subsubsection*{Antiferromagnetic resonance measurements}
Antiferromagnetic resonance measurements are performed in vacuum at cryogenic temperatures in an optical cryostat. 
The cryostat is positioned such that the sample space lies at the center of an electromagnet whose magnetic field lies in the plane of both the sample and the coplanar waveguide. 
The applied field was parallel to the MW field for all CrCl$_{3}$ measurements, and the applied field was perpendicular to the MW field for all CrSBr data (see Figs.~\ref{fig:S1} A and B). 

The primary independent variables in these measurements are the magnetic field strength and direction, the microwave power and frequency, and the temperature. 
Microwave frequency electric current is sent through the coplanar waveguide using a microwave generator. 
The microwave amplitude is modulated at low frequency to allow lock-in detection of the signal. 
The transmitted microwave amplitude is converted into a modulated DC voltage using a microwave diode, and the lockin detects this voltage using the modulation frequency as the reference frequency. 
At its resonance field and frequency, AFMR is inductively measured as a reduction of the transmitted microwave power. 

Simultaneously, green (532 nm) light generated by a laser is continuously applied to nanodiamonds on the sample surface containing NV$^-$ centers. 
The red photoluminescence (with green light filtered out passing $> 637$ nm) from the NV$^-$ centers is continuously collected by a photodiode whose output current is fed to a current-to-voltage preamplifier then fed into a second lockin amplifier referenced to the same modulation frequency. 
AFMR is optically measured as a reduction of emitted photoluminescence of the NV$^-$ centers.

\newpage



\begin{figure}
    \centering
    \includegraphics[width=0.75\textwidth]{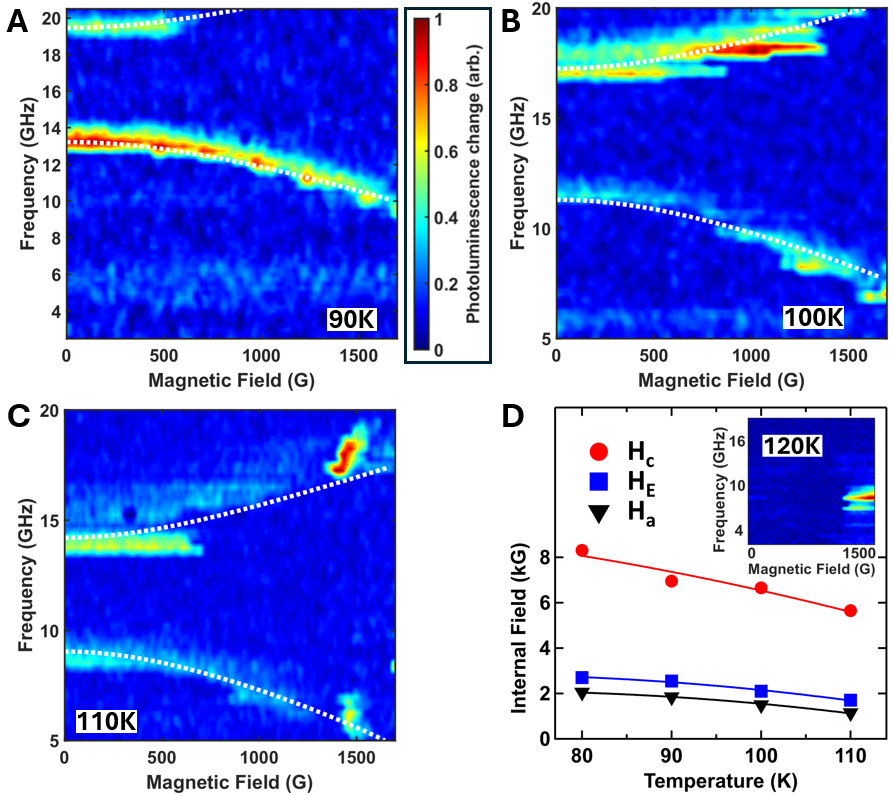}
    \caption{{\bf Determination of internal fields in CrSBr from optically detected AFMR spectra.} Panels (A) to (C) show the temperature evolution of the CrSBr dispersions of the lower frequency $\omega_-$  and the higher frequency $\omega_+$ modes both of which decrease with increasing temperature:
    panel (A) shows 90 K and panel (B) 100 K. In panel (C), taken at 110 K, both modes are visible along with a feature at 1500\,G that corresponds to either the spin-flop transition or a field-induced ferromagnetic mode.  
    Panel (D) summarizes the temperature dependencies of the internal fields $H_c$, $H_E$ and $H_a$; the white dashed curves in Figs.~\ref{fig:2}B and \ref{fig:2}C and in panels (A) to (C) show the mode dispersions predicted by these internal fields through Eq.~\ref{eq:CrSBr_modes}. Inset to panel (D):  No AFMR modes are observed at 120 K, but the feature near 1500 G intensifies.
    }
    \label{fig:3}
\end{figure}

\begin{figure}
    \centering
    \includegraphics[width=0.75\textwidth]{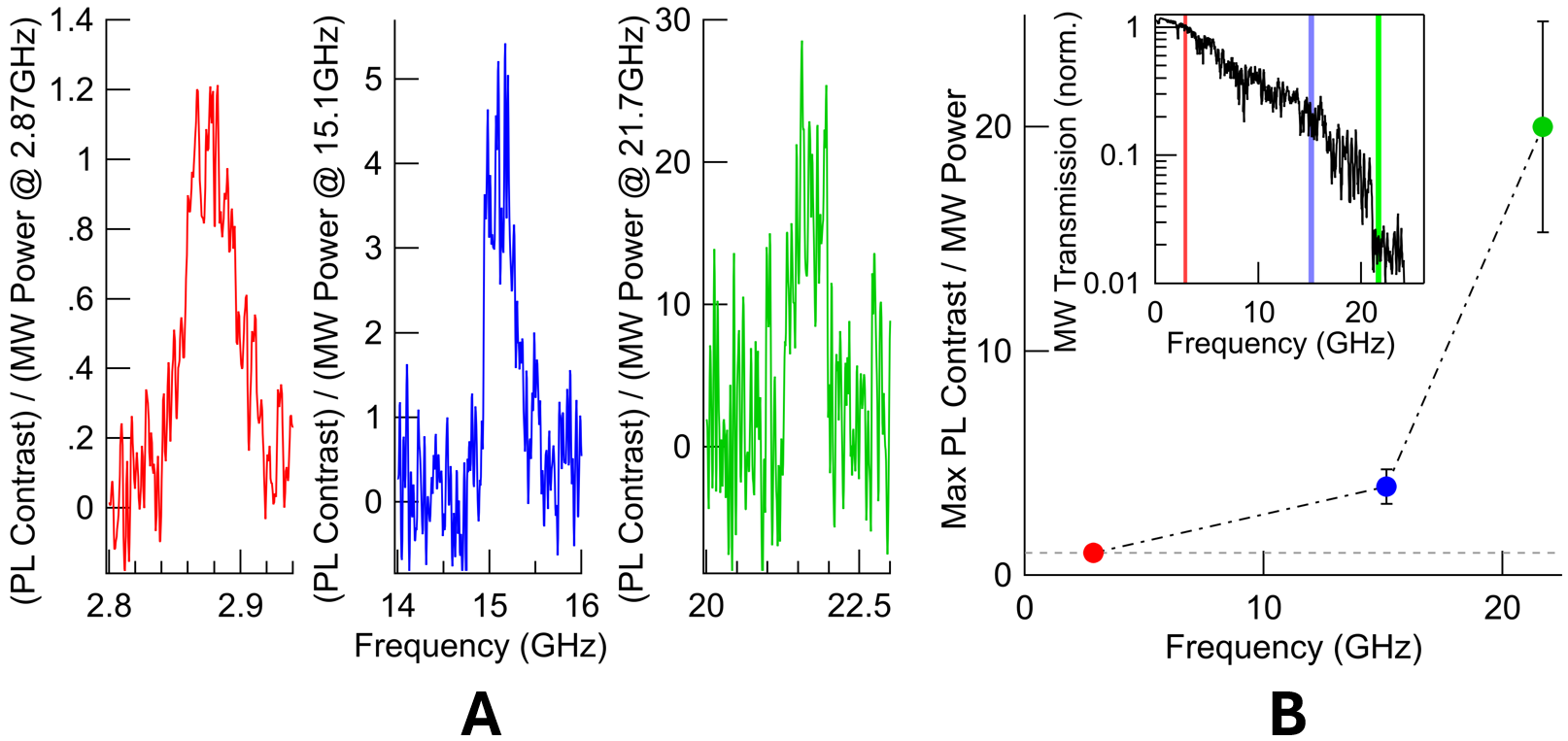}
    \caption{\textbf{Strength of optically detected magnetic resonance signals when normalized by microwave power.} Panel (A) shows photoluminescence contrast signals plotted on three different vertical scales to account for the substantial decrease of available microwave power at higher frequencies.  We compare the directly excited ESR (red), normalized to unity, with the 15.1\,GHz (blue) and 21.7\,GHz (green) CrSBr AFMR modes at 80K. The available microwave power, shown as an inset to panel B with vertical dashed lines at these frequencies, decreases by more than an order of magnitude between 15.1 and 21.7 GHz.  Panel (B) shows the frequency dependence of the size of the normalized signals obtained from the plots in panel (A). Correcting for this reduced power, the sizes of the signals of the low and high frequency modes exceed the direct NV$^{-}$ ESR by a factor of 4.5 and 20 respectively.}
    \label{fig:4}
\end{figure}


	


\clearpage 

%
%
\bibliography{QSAFdynamicsScienceSubmit} 

\begin{thebibliography}{10}
\providecommand{\url}[1]{\texttt{#1}}
\expandafter\ifx\csname urlstyle\endcsname\relax
  \providecommand{\doi}[1]{doi:\discretionary{}{}{}#1}\else
  \providecommand{\doi}{doi:\discretionary{}{}{}\begingroup \urlstyle{rm}\Url}\fi

\bibitem{BaltzRevModPhys}
V.~Baltz, A.~Manchon, M.~Tsoi, T.~Moriyama, T.~Ono, Y.~Tserkovnyak, Antiferromagnetic spintronics. \emph{Rev. Mod. Phys.} \textbf{90}, 015005 (2018).

\bibitem{SmejkalNatPhysTopoAFSpintronics}
L.~\v{S}mejkal, Y.~Mokrousov, B.~Yan, A.~H. MacDonald, Topological antiferromagnetic spintronics. \emph{Nat. Phys.} \textbf{14}, 242--251 (2018).

\bibitem{Huang1}
B.~Huang, G.~Clark, D.~R. Klein, R.~Cheng, E.~Navarro-Moratalla, K.~L. Seyler, D.~Zhong, E.~Schmidgall, M.~A. McGuire, D.~H. Cobden, D.~Xiao, W.~Yao, P.~Jarillo-Herrero, X.~Xu, Layer-dependent ferromagnetism in a van der {Waals} crystal down to the monolayer limit. \emph{Nature} \textbf{546}, 270--273 (2017).

\bibitem{Huang}
B.~Huang, G.~Clark, D.~R. Klein, D.~MacNeill, E.~Navarro-Moratalla, K.~L. Seyler, N.~Wilson, M.~A. McGuire, D.~H. Cobden, D.~Xiao, W.~Yao, P.~Jarillo-Herrero, X.~Xu, Electrical control of {2D} magnetism in bilayer {CrI}$_3$. \emph{Nat. Nanotechnol.} \textbf{13}, 544--548 (2018).

\bibitem{McGuire}
M.~A. McGuire, H.~Dixit, V.~R. Cooper, B.~C. Sales, Coupling of crystal structure and magnetism in the layered ferromagnetic insulator {CrI}$_3$. \emph{Chem. Mater.} \textbf{27}, 612--620 (2015).

\bibitem{lee2024spinorbitcouplingcontrolledtwodimensional}
I.~Lee, J.~Cen, O.~Molchanov, S.~Feng, W.~L. Huey, J.~van Tol, J.~E. Goldberger, N.~Trivedi, H.-Y. Kee, P.~C. Hammel, Spin-orbit coupling controlled two-dimensional magnetism in chromium trihalides  (2024), https://arxiv.org/abs/2405.16709.

\bibitem{Luqiao}
D.~MacNeill, J.~T. Hou, D.~R. Klein, P.~Zhang, P.~Jarillo-Herrero, L.~Liu, Gigahertz frequency antiferromagnetic resonance and strong magnon-magnon coupling in the layered crystal {CrCl}$_3$. \emph{Phys. Rev. Lett.} \textbf{123}, 047204 (2019).

\bibitem{DanRalphInductiveAFMRCrSBr}
T.~M.~J. Cham, S.~Karimeddiny, A.~H. Dismukes, X.~Roy, D.~Ralph, Y.~K. Luo, Anisotropic gigahertz antiferromagnetic resonance of the easy-axis van der {Waals} antiferromagnet {CrSBr}. \emph{Nano Lett.} \textbf{22}, 6716--6723 (2022).

\bibitem{CrSBrRoss}
J.~Klein, T.~Pham, J.~D. Thomsen, J.~B. Curtis, T.~Denneulin, M.~Lorke, M.~Florian, A.~Steinhoff, R.~A. Wiscons, J.~Luxa, Z.~Sofer, F.~Jahnke, P.~Narang, F.~M. Ross, Control of structure and spin texture in the van der {Waals} layered magnet {CrSBr}. \emph{Nat. Commun.} \textbf{13}, 5420 (2022).

\bibitem{CrSBrBingLv}
W.~Liu, X.~Guo, J.~Schwartz, H.~Xie, N.~U. Dhale, S.~H. Sung, A.~L.~N. Kondusamy, X.~Wang, H.~Zhao, D.~Berman, R.~Hovden, L.~Zhao, B.~Lv, A Three-Stage Magnetic Phase Transition Revealed in Ultrahigh-Quality van der {Waals} Bulk Magnet {CrSBr}. \emph{ACS Nano} \textbf{16}, 15917--15926 (2022).

\bibitem{BingLvNatComms}
X.~Guo, W.~Liu, J.~Schwartz, S.~H. Sung, D.~Zhang, M.~Shimizu, A.~L.~N. Kondusamy, L.~Li, K.~Sun, H.~Deng, H.~O. Jeschke, I.~I. Mazin, R.~Hovden, B.~Lv, L.~Zhao, Extraordinary phase transition revealed in a van der {Waals} antiferromagnet. \emph{Nat. Commun.} \textbf{15}, 6472 (2024).

\bibitem{Wolfe}
C.~S. Wolfe, V.~P. Bhallamudi, H.~L. Wang, C.~H. Du, S.~Manuilov, R.~M. Teeling-Smith, A.~J. Berger, R.~Adur, F.~Y. Yang, P.~C. Hammel, Off-resonant manipulation of spins in diamond via precessing magnetization of a proximal ferromagnet. \emph{Phys. Rev. B} \textbf{89}, 180406(R) (2014).

\bibitem{vanderSarYacobyProbingSpinWaves}
T.~van~der Sar, F.~Casola, R.~Walsworth, A.~Yacoby, Nanometre-scale probing of spin waves using single-electron spins. \emph{Nat. Commun.} \textbf{6}, 7886 (2015).

\bibitem{Chunhui_chemicalpotential}
C.~Du, T.~van~der Sar, T.~X. Zhou, P.~Upadhyaya, F.~Casola, H.~Zhang, M.~C. Onbasli, C.~A. Ross, R.~L. Walsworth, Y.~Tserkovnyak, A.~Yacoby, Control and local measurement of the spin chemical potential in a magnetic insulator. \emph{Science} \textbf{357}, 195--198 (2017).

\bibitem{Degen_AnnualReviewPChem}
R.~Schirhagl, K.~Chang, M.~Loretz, C.~L. Degen, Nitrogen-Vacancy Centers in Diamond: Nanoscale Sensors for Physics and Biology. \emph{Annu. Rev. Phys. Chem.} \textbf{65}, 83--105 (2014).

\bibitem{Wrachtrup_2006_Review}
F.~Jelezko, J.~Wrachtrup, Single defect centres in diamond: A review. \emph{Phys. Status Solidi A} \textbf{203}, 3207--3225 (2006).

\bibitem{JacquesAFMRnatcomm2020}
A.~Finco, A.~Haykal, R.~Tanos, F.~Fabre, S.~Chouaieb, W.~Akhtar, I.~Robert-Philip, W.~Legrand, F.~Ajejas, K.~Bouzehouane, N.~Reyren, T.~Devolder, J.-P. Adam, J.-V. Kim, V.~Cros, V.~Jacques, Imaging non-collinear antiferromagnetic textures via single spin relaxometry. \emph{Nat. Commun.} \textbf{12}, 767 (2021).

\bibitem{Vaidya_2020}
P.~Vaidya, S.~A. Morley, J.~van Tol, Y.~Liu, R.~Cheng, A.~Brataas, D.~Lederman, E.~del Barco, Subterahertz spin pumping from an insulating antiferromagnet. \emph{Science} \textbf{368}, 160 (2020).

\bibitem{RezendeAntiferromagneticMagnons}
S.~M. Rezende, A.~Azevedo, R.~L. Rodriguez-Suarez, Introduction to antiferromagnetic magnons. \emph{J. Appl. Phys.} \textbf{126}, 151101 (2019).

\bibitem{PurserNVseparation}
C.~M. Purser, V.~P. Bhallamudi, F.~Guo, M.~R. Page, Q.~Guo, G.~D. Fuchs, P.~C. Hammel, Spinwave detection by nitrogen-vacancy centers in diamond as a function of probe-sample separation. \emph{Appl. Phys. Lett.} \textbf{116}, 202401 (2020).

\bibitem{Maletinsky_Jacques}
L.~Rondin, J.-P. Tetienne, T.~Hingant, J.-F. Roch, P.~Maletinsky, V.~Jacques, Magnetometry with nitrogen-vacancy defects in diamond. \emph{Rep. Prog. Phys.} \textbf{77}, 056503 (2014).

\bibitem{FlebusTserkovnyak}
B.~Flebus, Y.~Tserkovnyak, Quantum-impurity relaxometry of magnetization dynamics. \emph{Phys. Rev. Lett.} \textbf{121}, 187204 (2018).

\bibitem{Brendan}
B.~A. McCullian, A.~M. Thabt, B.~A. Gray, A.~L. Melendez, M.~S. Wolf, V.~L. Safonov, D.~V. Pelekhov, V.~P. Bhallamudi, M.~R. Page, P.~C. Hammel, Broadband multi-magnon relaxometry using a quantum spin sensor for high frequency ferromagnetic dynamics sensing. \emph{Nat. Commun.} \textbf{11}, 5229 (2020).

\bibitem{Chunhui_hematite_ambient}
H.~Wang, S.~Zhang, N.~J. McLaughlin, B.~Flebus, M.~Huang, Y.~Xiao, C.~Liu, M.~Wu, E.~E. Fullerton, Y.~Tserkovnyak, C.~R. Du, Noninvasive measurements of spin transport properties of an antiferromagnetic insulator. \emph{Sci. Adv.} \textbf{8}, eabg8562 (2022).

\bibitem{CRD2018}
S.~Chatterjee, J.~F. Rodriguez-Nieva, E.~Demler, Diagnosing phases of magnetic insulators via noise magnetometry with spin qubits. \emph{Phys. Rev. B} \textbf{99}, 104425 (2019).

\bibitem{Machado2023}
F.~Machado, E.~A. Demler, N.~Y. Yao, S.~Chatterjee, Quantum Noise Spectroscopy of Dynamical Critical Phenomena. \emph{Phys. Rev. Lett.} \textbf{131}, 070801 (2023).

\bibitem{YJBaeNatureCoherentMagnonsCrSBr}
Y.~J. Bae, J.~Wang, A.~Scheie, J.~Xu, D.~G. Chica, G.~M. Diederich, J.~Cenker, M.~E. Ziebel, Y.~Bai, H.~Ren, C.~R. Dean, M.~Delor, X.~Xu, X.~Roy, A.~D. Kent, X.~Zhu, Exciton-coupled coherent magnons in a 2D semiconductor. \emph{Nature} \textbf{609}, 282 (2022).

\bibitem{Orenstein2024dipolar}
Y.~Sun, F.~Meng, C.~Lee, A.~Soll, H.~Zhang, R.~Ramesh, J.~Yao, Z.~Sofer, J.~Orenstein, Dipolar spin wave packet transport in a van der {Waals} antiferromagnet. \emph{Nat. Phys.} \textbf{20}, 794 (2024).

\bibitem{FoundationsMicrostripDesign_ch7}
T.~C. Edwards, M.~B. Steer, \emph{Foundations for Microstrip Circuit Design} (John Wiley \& Sons, Ltd), chap.~7, pp. 157--199 (2016).

\bibitem{HighFrequencyCPWchiang2023}
Y.-C. Chiang, H.-W. Tseng, C.-J. Yu, C.-Y. Lee, C.-C. Huang, C.-E. Ho, High-frequency signal transmission in a coplanar waveguide structure with different surface finishes. \emph{Thin Solid Films} \textbf{784}, 140079 (2023).

\bibitem{huang2023layer}
M.~Huang, J.~C. Green, J.~Zhou, V.~Williams, S.~Li, H.~Lu, D.~Djugba, H.~Wang, B.~Flebus, N.~Ni, C.~R. Du, Layer-dependent magnetism and spin fluctuations in atomically thin van der {Waals} magnet {CrPS$_4$}. \emph{Nano Lett.} \textbf{23}, 8099--8105 (2023).

\bibitem{Huang2023RevealingMicroscope}
M.~Huang, Z.~Sun, G.~Yan, H.~Xie, N.~Agarwal, G.~Ye, S.~H. Sung, H.~Lu, J.~Zhou, S.~Yan, S.~Tian, H.~Lei, R.~Hovden, R.~He, H.~Wang, L.~Zhao, C.~R. Du, {Revealing intrinsic domains and fluctuations of moir{\'{e}} magnetism by a wide-field quantum microscope}. \emph{Nat. Commun.} \textbf{14}, 5259 (2023).

\bibitem{Tschudin2023NanoscaleCrSBr}
M.~A. Tschudin, D.~A. Broadway, P.~Siegwolf, C.~Schrader, E.~J. Telford, B.~Gross, J.~Cox, A.~E.~E. Dubois, D.~G. Chica, R.~Rama-Eiroa, E.~J.~G. Santos, M.~Poggio, M.~E. Ziebel, C.~R. Dean, X.~Roy, P.~Maletinsky, Imaging nanomagnetism and magnetic phase transitions in atomically thin CrSBr. \emph{Nat. Commun.} \textbf{15}, 6005 (2024).

\bibitem{Xue2024SignaturesFerromagnet}
R.~Xue, N.~Maksimovic, P.~E. Dolgirev, L.-Q. Xia, R.~Kitagawa, A.~M{\"{u}}ller, F.~Machado, D.~R. Klein, D.~MacNeill, K.~Watanabe, T.~Taniguchi, P.~Jarillo-Herrero, M.~D. Lukin, E.~Demler, A.~Yacoby, {Signatures of magnon hydrodynamics in an atomically-thin ferromagnet}  (2024), https://arxiv.org/abs/arXiv:2403.01057.

\bibitem{ziffer2024quantum}
M.~E. Ziffer, F.~Machado, B.~Ursprung, A.~Lozovoi, A.~B. Tazi, Z.~Yuan, M.~E. Ziebel, T.~Delord, N.~Zeng, E.~Telford, D.~G. Chica, D.~W. deQuilettes, X.~Zhu, J.~C. Hone, K.~L. Shepard, X.~Roy, N.~P. de~Leon, E.~J. Davis, S.~Chatterjee, C.~A. Meriles, J.~S. Owen, P.~J. Schuck, A.~N. Pasupathy, Quantum Noise Spectroscopy of Criticality in an Atomically Thin Magnet  (2024), https://arxiv.org/abs/arXiv:2407.05614.

\bibitem{Das.hBN.PhysRevLett.133.166704}
S.~Das, A.~L. Melendez, I.-H. Kao, J.~A. Garc\'{\i}a-Monge, D.~Russell, J.~Li, K.~Watanabe, T.~Taniguchi, J.~H. Edgar, J.~Katoch, F.~Yang, P.~C. Hammel, S.~Singh, Quantum Sensing of Spin Dynamics Using Boron-Vacancy Centers in Hexagonal Boron Nitride. \emph{Phys. Rev. Lett.} \textbf{133}, 166704 (2024).

\bibitem{gong2024isotope}
R.~Gong, X.~Du, E.~Janzen, V.~Liu, Z.~Liu, G.~He, B.~Ye, T.~Li, N.~Y. Yao, J.~H. Edgar, E.~A. Henriksen, C.~Zu, Isotope engineering for spin defects in van der {Waals} materials. \emph{Nat. Commun.} \textbf{15}, 104 (2024).

\bibitem{PhysRevB.104.144416}
K.~Yang, G.~Wang, L.~Liu, D.~Lu, H.~Wu, Triaxial magnetic anisotropy in the two-dimensional ferromagnetic semiconductor CrSBr. \emph{Phys. Rev. B} \textbf{104}, 144416 (2021).

\bibitem{scheie2022spin}
A.~Scheie, M.~Ziebel, D.~G. Chica, Y.~J. Bae, X.~Wang, A.~I. Kolesnikov, X.~Zhu, X.~Roy, Spin waves and magnetic exchange Hamiltonian in CrSBr. \emph{Adv. Sci.} \textbf{9}, 2202467 (2022).

\bibitem{auerbach1998interacting}
A.~Auerbach, \emph{Interacting electrons and quantum magnetism} (Springer Science \& Business Media, New York, 1998) (1998).

\end{thebibliography}
\bibliographystyle{sciencemag}
\providecommand{\url}[1]{\texttt{#1}}
\expandafter\ifx\csname urlstyle\endcsname\relax
  \providecommand{\doi}[1]{doi:\discretionary{}{}{}#1}\else
  \providecommand{\doi}{doi:\discretionary{}{}{}\begingroup \urlstyle{rm}\Url}\fi

\section*{Acknowledgments}
\paragraph*{Funding:}
This research is primarily supported by the Center for Emergent Materials, a National Science Foundation (NSF) MRSEC, under award number DMR-2011876. S.C. acknowledges support from a PQI Community Collaboration Award. S.S. also acknowledges financial support from U.S. Office of Naval Research under Grant 298 N00014-23-1-2751 and NSF through grants DMR-2210510 and ECCS-2208057.  The work at UT Dallas is supported by the US Air Force Office of Scientific Research (AFOSR) Grant No.\ FA9550-19-1–0037, National Science Foundation (NSF) DMREF-2324033, and Office of Naval Research (ONR) Grant No.\ N00014-23-1–2020. Partial funding for the NanoSystems Laboratory shared facility used in this research is provided by the Center for Emergent Materials, an NSF MRSEC under award number DMR-2011876. 
\paragraph*{Author contributions:}
Conceptualization:  AM, SD, SS, PCH;   
Measurement and data collection:  AM, SD, FAR; 
Formal Analysis:  AM, SD, FAR, IK, SC, SS, PCH; 
Software:  AM, SD, IK, AW;
Materials:  AW, WL, BL, JG, IK, SS;
Project administration, Resources and Funding acquisition:  SS, PCH;	
Visualization: AM, FAR, AW, IK, SC, SS, PCH; 
Writing:  All authors.

\paragraph*{Competing interests:}  There are no competing interests to declare.
\paragraph*{Data and materials availability:}
All data are available in the manuscript or the supplementary materials.


\subsection*{Supplementary materials}
Supplementary Text\\ [-.6em]
Figs. S1 to S4\\


\newpage


\renewcommand{\thefigure}{S\arabic{figure}}
\renewcommand{\thetable}{S\arabic{table}}
\renewcommand{\theequation}{S\arabic{equation}}
\renewcommand{\thepage}{S\arabic{page}}
\setcounter{figure}{0}
\setcounter{table}{0}
\setcounter{equation}{0}
\setcounter{page}{1} 


\begin{center}
\subsection*{Supplementary Materials for\\ \scititle}

Alex~L.~Melendez$^{1\dagger}$,
    Shekhar~Das$^{1\dagger}$,
	Francisco~Ayala~Rodriguez$^{1}$,\\
    I-Hsuan~Kao$^{2}$,
    Wenhao~Liu$^{3}$,
	Archibald~J.~Williams$^{4}$,
    Bing~Lv$^{3}$,\\
    Joshua~Goldberger$^{4}$,
    Shubhayu~Chatterjee$^{2}$,
    Simranjeet~Singh$^{2}$,
    P.~Chris~Hammel$^{1\ast}$\\
	\small$^\ast$Corresponding author. Email: hammel.7@osu.edu\\
	\small$^\dagger$These authors contributed equally to this work.
\end{center}

\subsubsection*{This PDF file includes:}
Supplementary Text Sections: \\[-.4em]
 \emph{Aligning applied field with CrSBr easy-axis}  \\ [-.6em]
 \emph{Inductively Detected AFMR in CrCl${_3}$} \\ [-.6em]
 \emph{Dependence of AFMR on Laser Power}     \\ [-.6em]
 \emph{Magnon-scattering in CrSBr and Enhancement of NV$^{-}$ relaxation rate}  \\ [-.6em]
 \emph{Magnetization Configurations in an Applied Field} \\ 
 Figures S1 to S4\\


\newpage


\begin{figure}[h]
    \centering
    \includegraphics[width=0.75\textwidth]{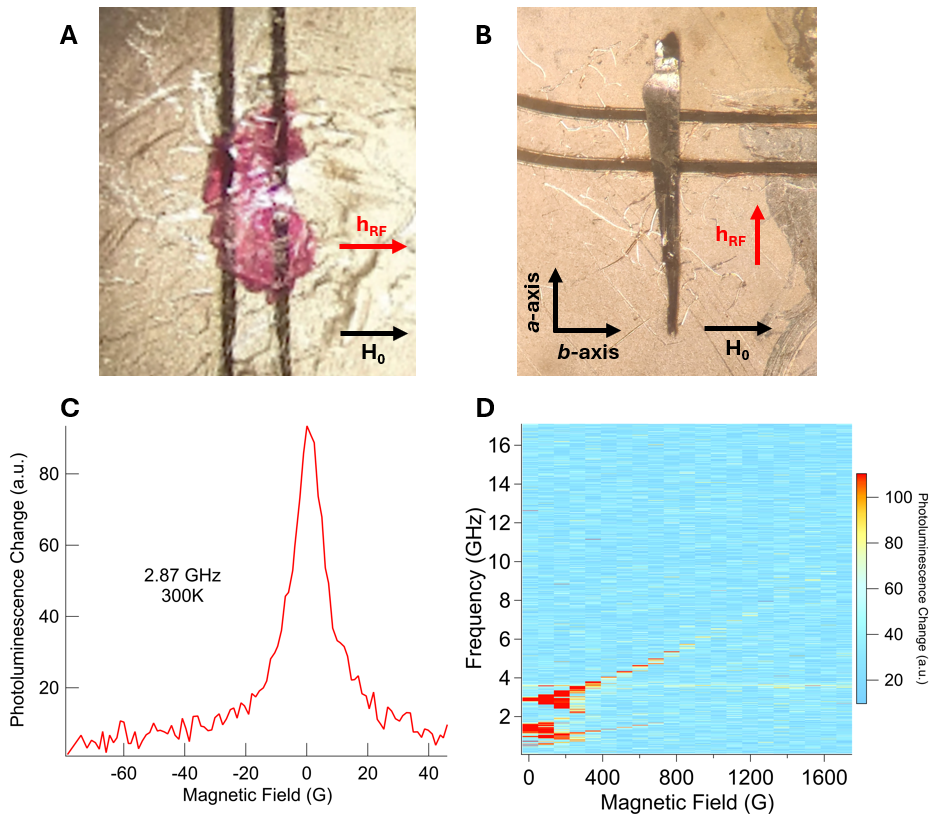}
    \caption{\textbf{Arrangement of samples with associated ESR spectra.}
    Optical images of CrCl$_{3}$, panel (A), and CrSBr, panel (B), on a microwave coplanar waveguide. The magnetic field 
    $\boldsymbol{H}_0$
    is applied along the left-right axis for both samples. 
    CrSBr is oriented such that the 
    easy-axis
    and the N\'eel vector 
    (crystallographic $b$-axis) are aligned along $\boldsymbol{H}_0$.
    Panel (C): Optically detected electron spin resonance (ODESR) of nanodiamonds deposited on CrCl$_{3}$ surface at room temperature at 2.87\,GHz. Panel (D): ODESR of nanodiamonds on CrSBr surface at room temperature. The data shown in panels (C) and (D) were obtained using a laser power of 5\,mW.}
    \label{fig:S1}
\end{figure}
\subsection*{Aligning applied field with CrSBr easy-axis}
The easy anisotropy axis of the CrSBr was determined via a combination of visual cues and the measured AFMR dispersion. When bulk CrSBr is viewed under an optical microscope and exfoliated, the difference between the two in-plane crystal axes (a-axis and b-axis) are visually evident due to the clear difference in their proclivity to cleave along the planes perpendicular to these directions. In particular, CrSBr tends to cleave along the plane perpendicular to the to the $b$-axis much more easily than cleaving along the plane perpendicular to the to the $a$-axis. We confirm this with measurements of the AFMR signal as a function of applied field and frequency in both geometries, that is, with the magnetic field parallel to and perpendicular to the $b$-axis. The dispersions obtained were consistent with the $b$-axis being the easy anisotropy axis and the $a$-axis being the intermediate anisotropy axis. This enabled us to align the applied magnetic field and the CPW center conductor such that both are parallel to the CrSBr easy-axis.

\subsection*{Inductively Detected AFMR in CrCl$_{3}$}

\begin{figure}[h]
    \centering
    \includegraphics[width=\textwidth]{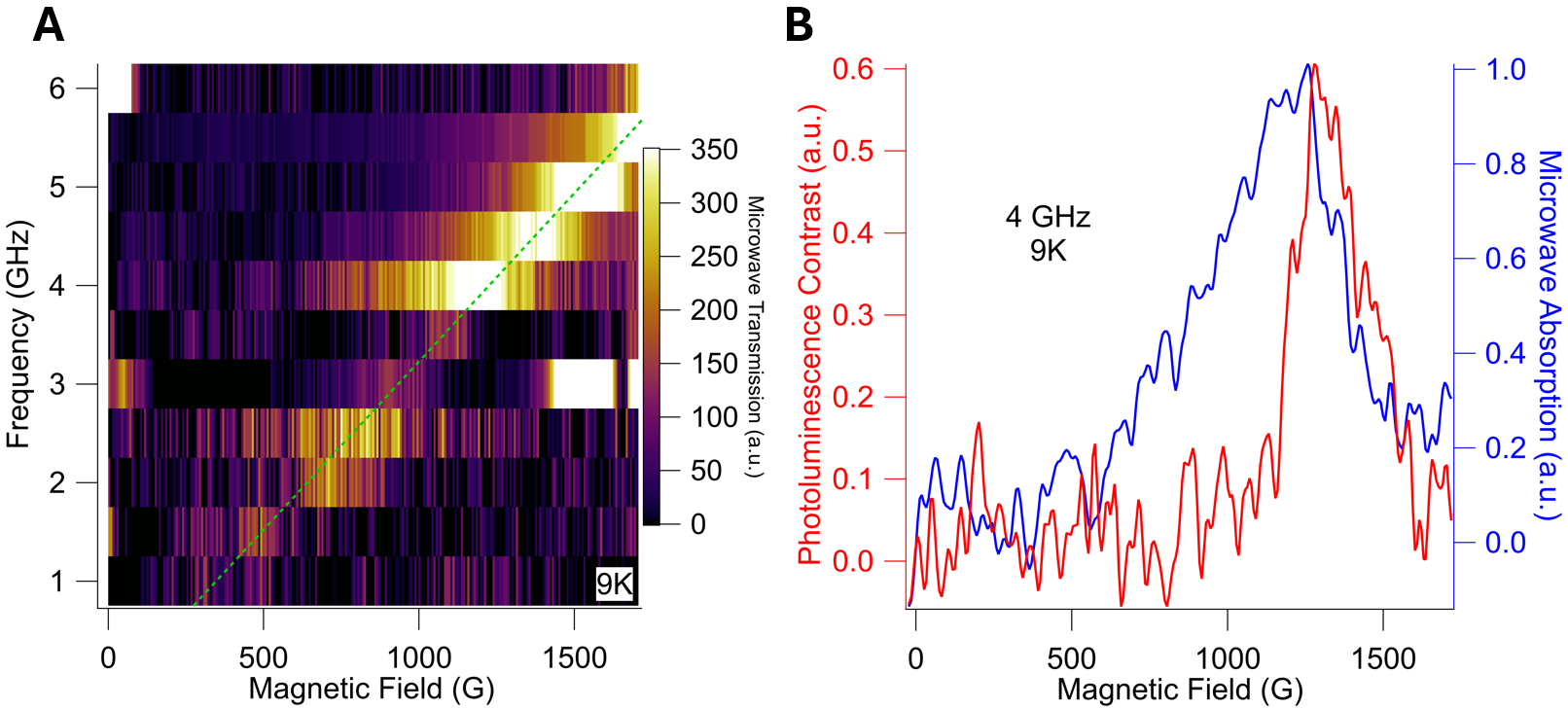}
    \caption{\textbf{Comparison of inductive and optically detected AFMR spectra.}
    Panel (A): Inductive AFMR (MW absorption) signal corresponding to the CrCl$_{3}$ optical data at 9K shown in Fig.~\ref{fig:2}A of the main paper. Dotted line is a fit to the MW absorption peaks as described in the main text following Eq.~\ref{modes} for the acoustic AFMR mode. Panel (B) compares optically detected [red] and inductively detected [blue] AFMR signals from CrCl$_{3}$ at 4\,GHz and 9\,K.}
    \label{fig:S2}
\end{figure}

Simultaneously with NV$^{-}$ photoluminescence measurements, AFMR was inductively measured by monitoring the reduction of microwave transmission ($S_{12}$) at 9K on CrCl$_{3}$, as shown in Fig.\,\ref{fig:S2}A. 
The frequency dependence of MW transmission between 0 and 12\,GHz required normalization of the inductive signal using MW transmission values obtained away from the resonance condition. 
The CrCl$_{3}$ optical and inductive signals at 9K and 4\,GHz are compared in Fig.\,\ref{fig:S2}B.

\subsection*{Dependence of AFMR on Laser Power}
\label{sec:PowerDependence}

\begin{figure}[h]
    \centering
    \includegraphics[width=0.5\textwidth]{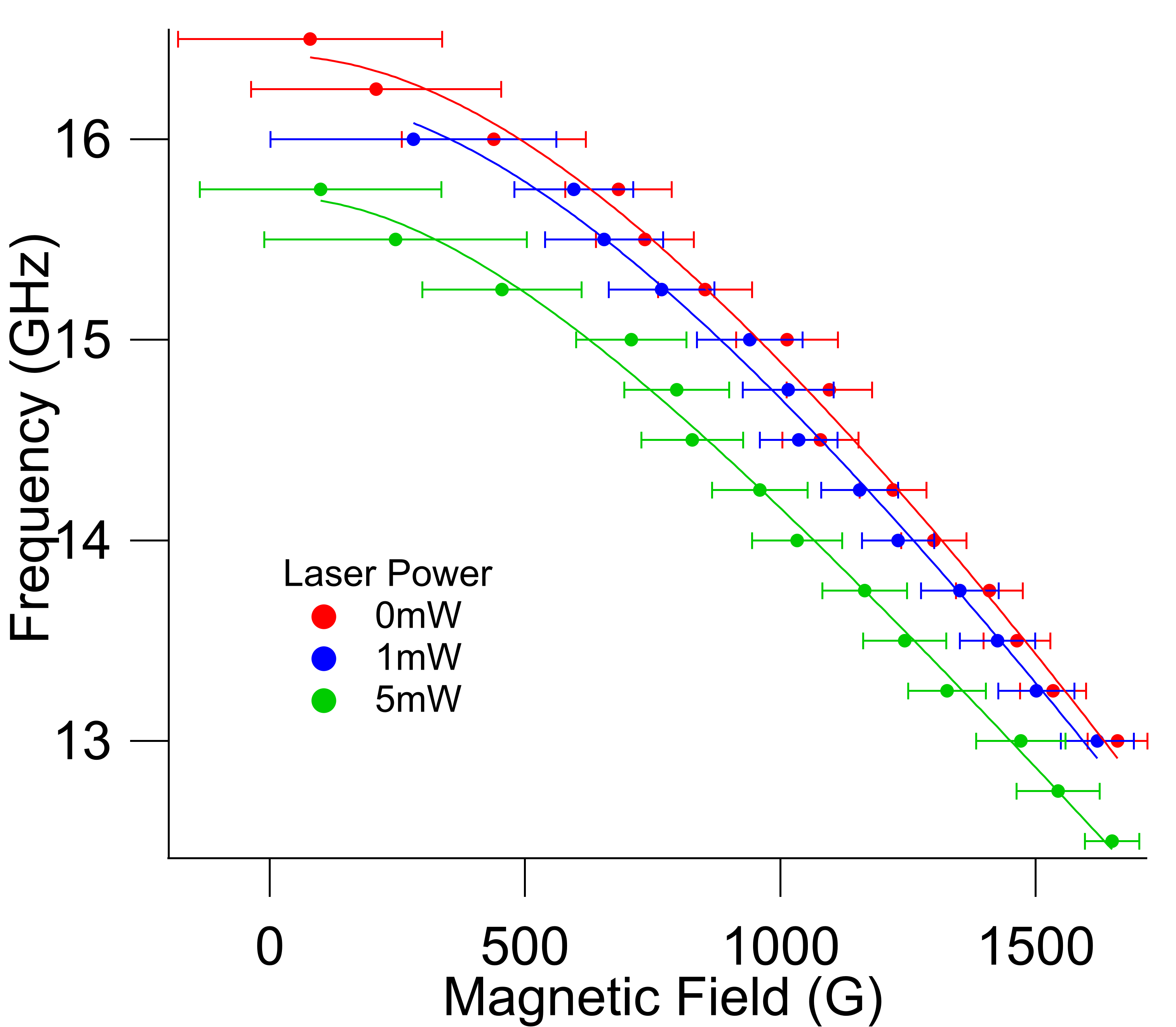}
    \caption{\textbf{Dependence of resonance frequencies on applied laser power in CrSBR.}
    Inductive AFMR signal of CrSBr low frequency mode at 80\,K, measured at three different laser powers. Solid curves show fits to Eq.\,\eqref{eq:CrSBr_modes}, main text. Error bars represent FWHM of Lorentzian fits.}
    \label{fig:S3}
\end{figure}

We measured the heating of the CrSBr sample by the laser (wavelength 532 nm) for the powers up to 5 mW used in our measurements. 
This done by comparing to inductive AFMR measurements performed at low (-10 dBm) MW power were performed in the presence of laser illumination at powers 0, 1mW and 5mW (Fig.~\ref{fig:S3}).  
To calibrate the temperature change associated with the change of the resonance field, the temperature dependence of the resonance field at 11\,GHz was taken from 90 to 107.5\,K which revealed a reduction of the resonance field of $(76.4 \pm 4.7)$\,G/K. 
Together with the change in the resonance field shown in Fig.~\ref{fig:S3} this yields an estimated temperature increase of $(0.8 \pm 0.5)$\,K at 1\,mW and $(3.3 \pm 0.5)$\,K at 5\,mW.

\def \B{{\boldsymbol{B}}}
\def \S{\boldsymbol{S}}
\def \k{{\boldsymbol{k}}}
\def \p{{\boldsymbol{p}}}
\def \q{{\boldsymbol{q}}}
\def \bd{\boldsymbol\delta}

\subsection*{Magnon-Scattering in CrSBr and Enhancement of NV$^{-}$ Relaxation Rate}
Here we work out the non-linear spin-wave Hamiltonian for CrSBr, and show that the quadratic terms lead to the magnon dispersions used in the main text, while the higher order terms lead to magnon-scattering. 
Next, we argue that when the zero-momentum magnon mode (AFMR mode) is pumped via resonant micro-wave excitation, it will decay via magnon-scattering into finite momentum modes, leading to spin transport. 
Finally, we show that the increased density of AFMR mode magnons will decrease the spin diffusion constant $D_s$ (assuming approximate number conservation in the steady state), leading to an increase in the NV$^{-}$ relaxation rate. 

The Hamiltonian for CrSBr is given by an in-plane ferro-magnetic coupling and inter-plane anti-ferromagnetic coupling between Cr$^{3+}$ spin $\S_{n,i}$ (where $n$ is the plane index and $i$ is the position of the spin in-plane), with a biaxial anisotropy that prefers alignment along the $b$ axis in the $ab$-plane, with the $a$ axis being the intermediate axis and the $c$ axis being the hard axis.
\begin{equation}
{\cal{H}} = - \sum_{n,i,\bd} J_{\bd} \, \S_{n,i} \cdot \S_{n,i + \bd} \, + \, J_E \sum_{n,i} \S_{n,i} \cdot \S_{n + 1,i} \, + \, \frac{D_c}{2} \sum_{n,i} \left(S^c_{n,i}\right)^2 \, + \, \frac{D_a}{2} \sum_{n,i} \left(S^a_{n,i}\right)^2 - g \mu_B \sum_{n,i} \B \cdot \S_{n,i}
\end{equation}
where $J_{\bd}$ refer to the in-plane coupling between spins at sites connected by $\bd$ ---the most significant ones are $J_2 = 3.88$ meV for $\bd = (\pm a \hat{a} \pm b \hat{b})/2$, $J_1 = 2.79$ meV for $\bd = \pm a \hat{a}$, and $J_3 = 2.12$ for $\bd = \pm b \hat{b}$ from first principles calculations in Ref. \cite{PhysRevB.104.144416}; slightly smaller values are obtained from neutron-scattering experiments in Ref.\cite{scheie2022spin}.
The anisotropies $D_a$ and $D_c$ and the inter-plane coupling $J_E$ are expected to be much smaller, of the order of 10-100 $\mu$eV.
They are related to the effective fields referenced in the main text by $H_{a(c)}=-D_{a(c)}S/g
\mu_B$ (similarly $H_E = - J_E S/g \mu_B$), where $g$ is the Land\'e g-factor, $\mu_B$ is the Bohr magneton and $S$ is the magnitude of the spin.
The small anisotropy terms $D_{a/c}$ break the continuous spin-rotation symmetry about the ordering direction $\hat{b}$ to discrete spin-rotation symmetries in $H$, indicating that the longitudinal spin-density will only be approximately (but not exactly) conserved by the system's dynamics.

The magnon dispersion for CrSBr has been derived using a Landau-Lifshitz approach \cite{DanRalphInductiveAFMRCrSBr}, but it does not readily provide the magnon-scattering terms. 
Therefore, to simultaneously derive both, we use non-linear spin-wave theory, \emph{i.e.}, we write the spin operator in terms of Holstein-Primakoff bosons and carry out a 1/S expansion \cite{auerbach1998interacting}.
For simplicity, we consider the case of a bilayer with $n = 1,2$, which is sufficient to capture the physics of interest. 
In the first layer, the spins are aligned along the easy axis, \emph{i.e.,}  $\langle \S_{1,i} \rangle = S \, \hat{b}$, while in the next layer, the spins are anti-aligned with the first layer, such that  $\langle \S_{2,i} \rangle = - S \, \hat{b}$, with $S = 3/2$.
We also consider $\B = B \hat{b}$ along the easy-plane, as in the experiment, and remain in the regime where $B$ is small enough to avoid a spin-flop transition.
Next, we define the spin-operators in terms of Holstein-Primakoff bosons $a$ and $b$ for the top and bottom layers as ($\hbar = 1$)
\begin{align}
S^b_{1,i} = S - a^\dagger_i a_i, ~ S^{+}_{1,i} = S^c_{1,i} + i S^a_{1,i} = (2S - a^\dagger_i a_i)^{1/2} a_i, ~ S^{-}_{1,i} = (S^+_{1,i})^\dagger = a^\dagger_i(2S - a^\dagger_i a_i)^{1/2}, \nonumber \\ 
S^b_{2,i} = -S + b^\dagger_i b_i, ~ 
S^{+}_{2,i} = S^c_{2,i} + i S^a_{2,i} = b_i^\dagger(2S - b^\dagger_i b_i)^{1/2},  ~ S^{-}_{1,i} = (S^+_{1,i})^\dagger = (2S - b^\dagger_i b_i)^{1/2} b_i
\end{align}
Expanding the Hamiltonian and carrying out an expansion order by order in 1/S, we arrive at the following Hamiltonian ($N$ = total number of sites in plane):
\begin{align}
{\cal{H}} &= {\cal{H}}^{(0)} + {\cal{H}}^{(2)} + {\cal{H}}^{(4)} + \ldots \nonumber \\
{\cal{H}}^{(0)} &= -2 N S^2 \left( \sum_{\bd} J_{\bd} \right) - J_E N S^2 \nonumber \\
{\cal{H}}^{(2)} & = S \left[ \sum_{i,\bd} J_{\bd} \left( a_i^\dagger a_i + a^\dagger_{i + \bd} a_{i + \bd} - a^\dagger_i a_{i + \bd} - a^\dagger_{i + \bd} a_i \right) + \sum_i \frac{(D_a - D_c)}{4}[a_i^2 - (a^\dagger_i)^2] \right. \nonumber \\
& ~~~~ \left. +\frac{(D_c + D_a)}{4} (a_i^\dagger a_i + a_i a_i^\dagger) \right] + ( a_i, \rightarrow b_i,  a_i^\dagger \rightarrow b_i^\dagger) + g \mu_B B \sum_i ( a^\dagger_i a_i - b_i^\dagger b_i) \nonumber \\
& ~~~~ + J_E S \left[ a_i^\dagger a_i + b_i^\dagger b_i + a_i^\dagger b_i^\dagger + a_i b_i \right]
\nonumber \\
{\cal{H}}^{(4)} & = \frac{1}{4} \sum_{i,\bd} J_{\bd} \left[ a^\dagger_i a_{i + \bd} a^\dagger_{i + \bd} a_i - a^\dagger_i (a_i)^2 a^\dagger_{i + \bd} - a_i (a^\dagger_{i + \bd})^2 a_{i + \bd} + \mbox{H.c.} \right] + \frac{1}{8} \sum_{i} (D_a - D_c)[a_i a^\dagger_i (a_i)^2 + (a^\dagger_i)^3 a_i] \nonumber \\
& ~~~~ + \frac{1}{8} \sum_i (D_c + D_a) [a_i (a_i^\dagger)^2 a_i + (a_i^\dagger)^2 (a_i)^2] + ( a_i, \rightarrow b_i,  a_i^\dagger \rightarrow b_i^\dagger) \nonumber \\
& ~~~~~ - \frac{J_E}{4} \sum_i \left[ a_i^\dagger a_i^2 b_i + a_i b_i^\dagger b_i^2 + (b_i^\dagger)^2 b_i a_i^\dagger + b_i^\dagger (a_i^\dagger)^2 a_i + 4 a_i^\dagger a_i b_i^\dagger b_i \right]
\end{align}

To find the magnon spectrum, we consider the quadratic part ${\cal{H}}^{(2)}$, which is linear in $S$ (except the term proportional to the field) and can be conveniently written in terms of a Nambu spinor $\Psi_\k$ in Fourier space.
\begin{align}
{\cal{H}}^{(2)} &= \sum_\k \Psi^\dagger_\k h^{(2)}(\k) \Psi_\k, \text{ where } \Psi_\k = \begin{pmatrix}
   a_\k &
   b_\k &
   a^\dagger_{-\k} &
   b^\dagger_{-\k}
\end{pmatrix}^T,  \nonumber \\
h^{(2)}(\k) &= \frac{1}{2} \begin{pmatrix}
 f_\k + g \mu_B B & 0 & \frac{(D_c - D_a)S}{2} & J_E S  \\
 0 & f_\k - g \mu_B B & J_E S & \frac{(D_c - D_a)S}{2} \\
  \frac{(D_c - D_a)S}{2} & J_E S & f_\k + g \mu_B B & 0 \\
  J_E S & \frac{(D_c - D_a)S}{2} & 0 & f_\k - g \mu_B B
\end{pmatrix}
\end{align}
and we have defined 
\begin{align}
a_\k & = \frac{1}{\sqrt{N}} \sum_i e^{i \k \cdot \mathbf{r}_i} a_i \, , \, b_\k  = \frac{1}{\sqrt{N}} \sum_i e^{i \k \cdot \mathbf{r}_i} b_i \, , \text{ and } \nonumber \\
f_\k &=  S \left[ \gamma_\k  + \frac{(D_c + D_a)}{2} + J_E \right] \text{ with } \gamma_\k = 2\sum_{\bd} J_{\bd} [1 - \cos(\k \cdot \bd)]. 
\end{align}
Finding the dispersion of the two magnon branches $\omega_{\mu}(\k)$ (labeled by $\mu = \pm$) then amounts to diagonalizing the following dynamical matrix (that preserves bosonic commutation relations). 
\begin{align}
 D(\k) = M {\cal{H}}^{(2)}(\k), \text{ where } M = \begin{pmatrix}
    1 & 0 & 0 & 0 \\
    0 & 1 & 0 & 0 \\
    0 & 0 & -1 & 0 \\
    0 & 0 & 0 & -1 \\
 \end{pmatrix}
\end{align}
On doing so, we may write the quadratic part in terms of the magnon creation operators $\alpha^\dagger_{\mu, \k} $ as
\begin{align}
{\cal{H}}^{(2)} = \sum_{\k, \mu = \pm } \omega_{\mu}(\k) \left[ \alpha^\dagger_{\mu, \k} \alpha_{\mu, \k} + \frac{1}{2} \right], \text{ where }
\end{align}
\begin{align}
\omega_{\pm}^2(\k) &= (g \mu_B B)^2 + S^2 \left[(D_a + D_c) J_E + D_a D_c + (D_a + D_c + 2 J_E) \gamma_\k + \gamma_\k^2 \right] \pm S \bigg[ (J_E S)^2(D_a - D_c)^2  \nonumber \\
& ~~~~~ + (g \mu_B B)^2 (D_a + D_c)(D_a + D_c + 4 J_E)   + 4 (g \mu_B B)^2 \{ (D_a + D_c + 2 J_E) \gamma_\k + \gamma_\k^2 \} \bigg]^{1/2}
\label{eq:dispk}
\end{align}
We note that setting $\k = 0$ in Eq.~\eqref{eq:dispk} implies $\gamma_\k = 0$, and this allows us to recover the dispersions of the two uniform AFMR modes $\omega_\pm$ discussed in the main text. 

Next, we turn to the magnon-scattering terms, which may be re-written in terms of the $\alpha_\mu$ bosons, which reads (suppressing the $\mu$ index for clarity) 
\begin{align}
{\cal{H}}^{(4)} &=  \frac{1}{N} \sum_{\k, \q, \p} V^{(4)}_{\k,\q,\p} \alpha^\dagger_{\k} \alpha^\dagger_{\q} \alpha_{\p} \alpha_{\k + \q - \p} +  \bar{V}^{(4)}_{\k,\q, \p} \alpha^\dagger_{\k} \alpha_{\q} \alpha_{\p} \alpha_{\k - \q - \p} + \mbox{H.c.}. 
\end{align}
The magnon-scattering terms can be grouped into magnon number-conserving terms like $V^{(4)}$, and number-non-conserving terms like $\bar{V}^{(4)}$.
The result of the presence of these terms would be to redistribute the population of the magnons when the $\k = 0$ mode (AFMR mode) is excessively populated to reach a steady state.
The transient dynamics is a complicated kinetic problem, so we proceed by making the physically reasonable assumption that a non-equilibrium steady state is reached with a finite density of non-zero momentum magnons.  
The collective dynamics of these magnons is responsible for spin diffusion, and enhances the relaxation-rate of the NV$^{-}$ centers placed proximate to the sample.

For simplicity, let us consider a single NV$^{-}$ center placed at a distance $d$ from the CrSBr sample. 
The relaxation rate of the NV$^{-}$ is mainly due to magnetic noise caused by longitudinal collective modes, as the magnon gaps $\omega_\pm$ are much larger than the NV$^{-}$ splitting $\Omega$. 
Therefore, we restrict ourselves to considering magnetic noise at the NV$^{-}$ from such collective magnon modes, which in turn determine the longitudinal spin-correlations $C_{\parallel}(\k,\Omega)$ in the steady state. 
The corresponding relaxation rate (due to a CrSBr sample at distance $d$) is given by 
\begin{equation}
\Gamma_1(\Omega) \propto \int d^2k \, k^2 e^{-2 k d} \, C_\parallel(\k, \Omega).
\end{equation}
The net relaxation rate may be obtained via adding the relaxation rate due to each layer, assuming the layers are weakly correlated. 

If the longitudinal spin-density (set by the magnon-number) was conserved and the system was in equilibrium, we would expect $ C_\parallel(\k, \Omega)$ to take a diffusive form at the longest length-scales and time-scales.
While neither is precisely true for our problem, the energy-scale of non-conservation (set by $D_{a/c}$) is a thousand times smaller relative to the dominant energy-scales $J_{1/2/3}$, and we also assume that the sample is near equilibrium, so that the spin-density is approximately conserved. 
Therefore, we may write (via the fluctuation-dissipation theorem)
\begin{equation}
 C_\parallel(\k, \Omega) = \frac{2 k_B T}{\Omega} \text{Im}[\chi(\k,\Omega)], \text{ where } \chi(\k,\Omega) = \frac{\chi(D_s k^2 + \gamma_s)}{- i \Omega + D_s k^2 + \gamma_s}
\end{equation}
is the dynamical susceptibility, $D_s$ is the spin diffusion constant, and $\gamma_s$ captures the deviation from spin-density conservation. 
Since the NV$^{-}$ frequency $\Omega$ is the lowest energy-scale in the problem, we can set it to zero for analytical convenience.
We find that the relaxation rate is given by
\begin{align}
\Gamma_1 \propto \int d^2k \, k^2 e^{- 2 k d} C_\parallel(\k, \Omega \to 0)= \int d^2k \, k^2 e^{- 2 k d} \frac{2 k_B T \chi}{D_s k^2 + \gamma_s} \xrightarrow{D_s \gg \gamma_s d^2} \frac{2 k_B T \chi}{d^2 D_s}
\end{align}
The last approximation is a good approximation if the net spin-density (along $b$ axis) is conserved, i.e., the  magnon number is conserved, i.e., $\gamma_s \to 0$.

All that remains to be determined is the functional dependence of $\chi$ and $D_s$ on the magnon density $n$.
To derive this, we first note that each AFMR mode has a finite longitudinal  magnetization density (approximately, $\mathbf{m} \propto \chi(\mathbf{n} \times \dot{\mathbf{n}})$ where $\mathbf{n}$ is the Neel vector), as shown in Fig. 2(b),(c) in the main text. 
This spin density $\mathbf{m} \cdot \hat{b}$ is directly proportional to the number density of magnons $n$, so we need to essentially determine the spin-susceptibility and diffusivity due to magnon dynamics.
It is convenient to use the Einstein relation that the magnon conductivity $\sigma_s = \chi D_s$, then we find that $\Gamma_1 \sim \sigma_s D_s^{-2}$.
Now, we note that within a semi-classical picture of magnon conduction, the magnon conductivity $\sigma_s \propto n \tau$, where $n$ is the density and $\tau$ is the scattering time.
If magnon-scattering dominates momentum relaxation, then $\tau \propto 1/n$, implying that $\sigma_s$ is independent of the density.
Physically, the effect of a larger number of carriers is canceled by the fact that the carriers also collide more frequently. 
On the other hand, the diffusion constant $D_s \sim \langle v^2 \rangle \tau \propto \langle v^2 \rangle/n$, where $\langle v^2 \rangle$ is the mean-squared velocity that is set by the temperature, so the diffusion constant decreases with increasing density $n$. 
Combining these two results, we find the density dependence of the NV$^{-}$ relaxation rate to be $\Gamma_1 \sim \sigma_s D_s^{-2} \sim n^2$.
Thus, we have shown that an increased number density of magnons, created by pumping, can cause faster relaxation of a proximate NV$^{-}$ center. 

\subsection*{Magnetization Configurations in an Applied Field}
\label{sec:MagnetizationGeometry}

\begin{figure}[h]
    \centering
    \includegraphics[width=0.8\textwidth]{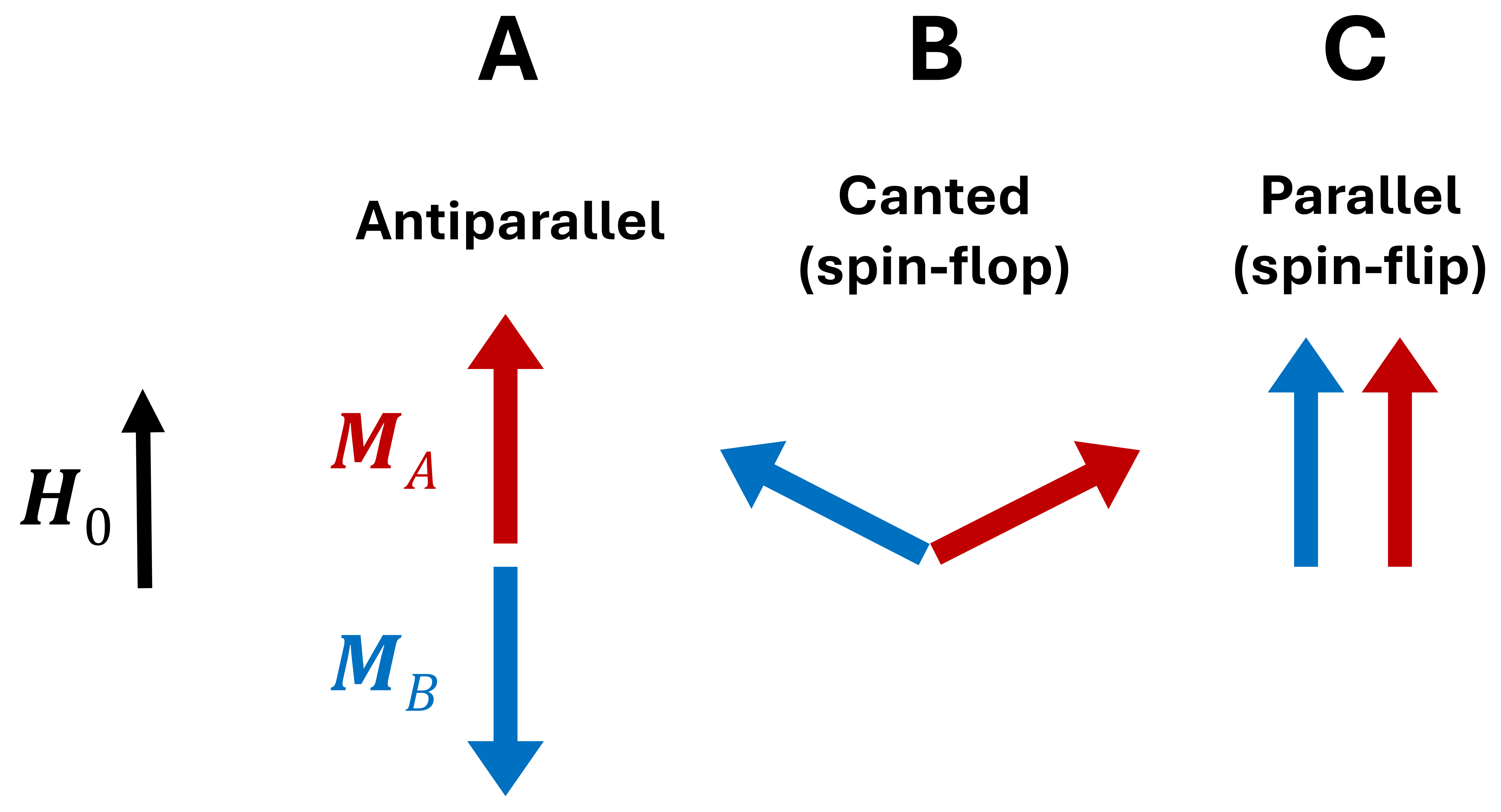}
    \caption{
    \textbf{Magnetization configurations for an antiferromagnet in an applied magnetic field.}
    Three configurations of the antiferromagnetic sublattice magnetizations for different strengths of field applied parallel to the anisotropy easy axis. Panel (A): From zero to low applied field $H_0$, the sublattice magnetizations remain aligned with the (here vertical) easy-axis and mutually antiparallel. Panel (B): At intermediate fields, the sublattice magnetizations lie in a canted configuration. Panel (C) At higher fields, the sublattice magnetizations lie parallel to their counterparts and to the applied field.
    }
    \label{fig:S4}
\end{figure}

The two configurations of the antiferromagnetic sublattice magnetizations mentioned in the main text are those of the antiparallel (``below spin-flop'') and canted (``spin-flop'') geometries (see Fig.\,\ref{fig:S4}). 
In zero applied field, the equilibrium state of both antiferromagnets is the antiparallel state, with the sublattice magnetizations lying along an easy-axis as in Fig.\,\ref{fig:S4}A. 
In a magnetic field applied perpendicular to this axis the spins will  cant in the direction of the field at arbitrarily small field, and this canting will increase with field asymptotically approaching the parallel or ``spin-flip'' state as shown in Fig.~\ref{fig:S4}B. 

The behavior is different when the magnetic field is applied parallel to the easy axis. In this case the sublattice magnetizations will remain antiparallel until the field reaches the spin-flop field $H_\text{SF} \approx \sqrt{2 H_E H_A}$, where $H_E$ is the exchange field and $H_A$ is the easy-axis anisotropy. 
Above $H_\text{SF}$ the sublattice magnetizations abruptly reorient to the canted configuration and increasingly align with the applied field as the field increases. 
At an even higher field, the increasing contribution of the anisotropy causes the sublattice magnetizations to undergo another abrupt reorientation to fully align with the field in the parallel configuration shown in Fig.~\ref{fig:S4}C.

In the case of CrCl$_3$ AFMR, because the in-plane anisotropy is negligible, the sublattice magnetizations reorient to the canted configuration of Fig\,\ref{fig:S4}B in the presence of even very small applied field. Since there is no in-plane easy axis, the sublattice magnetizations remain in this configuration at any finite field in our field range. 

In contrast, for CrSBr the field is applied along the in-plane easy axis, and thus the sublattice magnetizations remain in the antiparallel configuration shown in Fig.~\ref{fig:S4}A for all measurements, except for that shown in the inset to Fig.~\ref{fig:3}D where around 1500\,G and 120\,K we observe what may be antiferromagnetic resonance in the spin flop configuration of Fig.\,\ref{fig:S4}B.  
In all cases, antiferromagnetic resonance consists of small-angle precession of the sublattice magnetizations about these equilibrium directions.





\end{document}